\documentclass[12pt,preprint]{aastex}

\usepackage{natbib}
\usepackage{amsmath,amsfonts,amssymb}
\usepackage{graphicx}

\slugcomment{To be Submitted to Astrophysical Journal}

\shorttitle{HD 80606b}
\shortauthors{Langton \& Laughlin}

\begin{document}

\title{A New Atmospheric Model for HD 189733 b}

\author{Jonathan Langton}
\affil{Department of Astronomy and Astrophysics\\ University of California at Santa Cruz\\
    Santa Cruz, CA 95064}
\email{jlangton@ucsc.edu}

\and
\author{Gregory Laughlin}
\affil{UCO/Lick Observatory\\ 
Department of Astronomy and Astrophysics\\
University of California at Santa Cruz\\
Santa Cruz, CA 95064}
\email{laughlin@ucolick.org}

\begin{abstract}
We have developed a new two-dimensional hydrostatically-balanced isobaric hydrodynamic model for use in simulation of exoplanetary atmospheres.  We apply this model to the infrared photosphere of the hot Jupiter HD 189733 b, for which an excellent 8-$\mu$m light curve has been obtained.  For reasonable parameter choices, the results of our model are consistent with these observations.  In our simulations, strongly turbulent supersonic flow develops, with wind speeds of approximately 5~km~s$^{-1}$.  This flow geometry causes chaotic variation of the temperature distribution, leading to observable variations in the light curve from one orbit to the next.
\end{abstract}


\section{Introduction}
Since the earliest dynamical models of strongly irradiated exoplanetary atmospheres were published \citep{sho02,cho03}, it has been apparent that the expected large temperature gradients and the resulting high wind speeds could have such a large impact on these planets' appearance that the effects could be observable even at a distance of dozens of parsecs.  As the models have grown increasingly sophisticated over the past six years, research has progressed along two distinct lines.  \citet{sho02}, \citet{coo05}, \citet{dob08} and \citet{sho08} have produced three-dimensional models, in an effort to simulate the greatest possible range of relevant physical processes.  Because of the high computational costs of such models, they have been run at comparatively low resolution.  In contrast to this approach, \citet{cho03}, \citet{lan07}, \citet{lan08}, and \citet{lan08b} have chosen to employ two-dimensional models which can be run at higher resolutions.  On highly irradiated planets, the radiative zone is believed to extend deep into the atmosphere, to a pressure depth of hundreds of bars \citep{coo05, iro05, sho08}.  The flow is therefore expected to be strongly stratified, with vertical motion comparatively unimportant.  The assumption inherent in a two-dimensional model is that the motion at small scales which can be captured due to the finer grid spacing is more important than the vertical flow which a two-dimensional model must neglect.  Nevertheless, a serious concern for two-dimensional models is the possible existence of crucial three-dimensional processes which do not require strong vertical motion in order to become significant.

Furthermore, previous attempts to develop two-dimensional models have been hampered by questions regarding the validity of the physics involved.  \citet{lan07} employ a shallow-water model which is at best a first-order approximation to realistic atmospheric dynamics.  While the barotropic-equivalent model used by \citet{cho03}, formally very similar to the shallow-water equations, is on a firmer physical footing, there is some indication that the model becomes numerically unstable at high wind speeds \citep{rau07}.  In an attempt to achieve greater realism than is possible using a shallow-water model, \citet{lan08} developed a model employing fully-compressible two-dimensional hydrodynamics.  As we will show in \S2, however, such models produce features which necessarily violate hydrostatic balance.  In this paper, then, we present a two-dimensional model which must maintain hydrostatic equilibrium.
It is our hope that this new model will be able to approach the rigor of a three-dimensional model, while maintaining the adaptability and speed of a two-dimensional model.

While these modeling efforts have been underway since 2002, the ability to constrain these models using observations is a more recent development.  Of particular interest, \citet{knu07} have produced a map of the planet's longitudinal temperature variation based on their \textit{Spitzer} observations of its flux in the 8-$\mu$m band.  From the depth of the secondary eclipse, they find a hemispherically-averaged day-side brightness temperature of $1205.1 \pm 9.3$ K, while they estimate a cooler hemispherically-averaged night-side brightness temperature of $973 \pm 33$ K, based on the flux curve.  Interestingly, the hottest part of the planet is \emph{not} directly beneath the star: the temperature maximum is offset some $30^\circ$ east of the substellar point.  Perhaps even more interestingly, the temperature minimum is offset $30^\circ$ \emph{west} of the antistellar point.  It is clear, then, the temperature distribution is strongly influenced by planetary winds; at first glance it would appear that the flow is eastward on the day-side, while westward on the night-side.  To date, models have not reproduced this flow geometry \citep{cho03, coo05, lan07, cho08, dob08, lan08, sho08}.  Additionally, these observations do not seem to be generally applicable to hot Jupiters.  \citet{har06}, for example, observed the 24-$\mu$m flux of the giant exoplanet $\upsilon$ Andromedae b, finding a much larger temperature contrast between the illuminated and dark hemispheres.  Additionally, they found a slight \emph{westward} displacement of the hot spot from the substellar point, although their observations did not exclude the possibility that there was no offset at all.  The mid-infrared observations of HD 179949 b by \citet{cow07} also appear to be consistent with zero phase offset.  It is not clear to what extent these discrepancies result from different dynamics on the planets themselves, and to what extent they are caused by the different pressure depths under observation.

In other respects, HD 189733 b remains a fairly typical hot Jupiter; its period, however, is shorter than most, with $P=2.218573 \pm 0.000020$ d.  The transit has allowed determinations of several other parameters: $R=1.154 \pm 0.032 R_J$, $i=85.79^\circ \pm 0.24^\circ$ \citep{bak06}, $a=0.0313 \pm 0.0004$, and $M=1.15 \pm 0.04 M_J$ \citep{bou05}.  The orbit is assumed to be circular.

HD 189733 b has not received the attention from modelers which has been bestowed on its more famous cousin, HD 209458 b.  Nevertheless, the well-resolved flux curve produced by \citet{knu07} has made it a most interesting target for simulation, and \citet{sho08} apply their atmospheric model to HD 189733 b, with mixed success.  They recover the correct eastward phase offset for the hot spot, but are unable to produce a flow pattern which causes the observed westward offset for the cold spot from the antistellar point.  Their flow patterns in the upper atmosphere are in general supersonic, with very low pressures characterized by longitudinal and latitudinal flow from the day side towards the night side, while deeper layers are dominated by a supersonic eastward jet at the equator.  Interestingly, despite the high wind speeds ($|\mathbf{v}| \gtrsim 3$ km/s) and the resulting large wind shear, no turbulence develops in their simulations.  As we shall see, this does not match the results of the two-dimensional model presented in this paper.  It is possible that their relatively low horizontal resolution (144x90) combined with the finite-difference differentiation employed by the ARIES/GEOS dynamical core used in their model conspire to produce sufficient numerical dissipation to prevent the development of turbulence; it is also possible (and possibly more likely) that the development of turbulence is prevented by three-dimensional effects which our two-dimensional model is unable to capture.  An analytical study to constrain the conditions under which turbulence is expected to arise would surely be profitable; however, this, and indeed, the more general question of the conditions necessary for planetary-scale turbulent flow, must remain a topic for future investigation.

In any case, it is clear that the \citet{knu07} time-series is of sufficiently high quality to provide an excellent benchmark for both existing and future simulations of exoplanetary atmospheres.  The ARIES/GEOS core used in \citet{coo05} and \citet{sho08} has been applied to the terrestrial atmospheres of Earth and Mars \citep{sho08}, and the equivalent barotropic formulation of \citet{cho03} -- also used in \citet{rau07} -- has enjoyed some success in reproducing the primary features of Jupiter's dynamics.  However, the conditions on many extrasolar planets are so utterly unlike anything seen in our solar system that it is advisable to include data from exoplanet observations when testing the models.

This paper is organized as follows: In \S 2, we derive in some detail the hydrodynamical core of our model, as well as providing a treatment of the radiative forcing scheme.  In \S 3, we apply our model to the atmosphere of HD 189733 b, comparing our results to those of \citet{sho08} and to the data obtained by \citet{knu07}.  We conclude in \S4.

\section{Numerical Model}

We begin with the equations of motion in three dimensions for an irradiated, hydrostatically-balanced ideal gas in a rotating reference frame, with an isobaric vertical coordinate: \citep{sal96}
\begin{align}
\frac{\partial \Phi}{\partial p} &= -\frac{RT}{p} \label{p31}\\
\frac{\partial w_p}{\partial p} &= -\nabla \cdot \mathbf{v} \label{p32}\\
\frac{\partial \mathbf{v}}{\partial t} &= -\mathbf{v} \cdot \nabla \mathbf{v} - w_p \frac{\partial \mathbf{v}}{\partial p} - \nabla{\Phi}-2\Omega_{\rm{rot}} \sin \theta (\hat{n} \times \mathbf{v}) \label{p33}\\
\frac{\partial T}{\partial t} &= -\mathbf{v} \cdot \nabla T  - w_p \left( \frac{\partial T}{\partial p}-\frac{\kappa T}{p} \right) + f_{\rm{rad}}, \label{p34}
\end{align}
where $\Phi=gz$ is the geopotential at constant pressure, $w_p = dp/dt$ is the ``velocity'' in pressure coordinates (while $w_p$ does not have units of velocity, the quantity $w_p \partial/\partial p$ does have the correct units of advection), and $\kappa = 1-1/\gamma = 2/7$ for an ideal diatomic gas.  It is also important to note that $\mathbf{v}$ describes \emph{only} the horizontal components of the flow; vertical motion, associated with the upward or downward motion of an isobar, necessary to maintain hydrostatic equilibrium, is encapsulated in $w_p$.  

To adapt this three-dimensional system to two dimensions, it is necessary to assume a particular vertical structure for $T$ and $\mathbf{v}$.  We assume that the temperature does not vary with pressure; this is a reasonable first approximation, and any temperature variation is not expected to cause significant deviations in the geopotential.  To see this, consider an atmosphere where the temperature can vary (weakly) with pressure:  $T(p) = T_0 + \Gamma p$.  (This is, of course, a gross simplification, as a temperature that was a function solely of pressure would not give rise to any interesting dynamics if one is looking at layers of constant pressure!)  The geopotential is found by integrating equation \ref{p31} from some pressure $p_b$ at the boundary to $p$, the pressure of the single layer being simulated in our two-dimensional model.  We further assume that the isobar at $p_b$ is at constant geopotential, so that we can set $\Phi(p_b) = 0$.  Then
\begin{align*}
\Phi(p) &= -R \int_{p_b}^p \frac{T dp'}{p'}\\
\Phi(p) &= -RT_0 \int_{p_b}^p \frac{dp'}{p'}-R\int_{p_b}^p \Gamma \,dp'\\
\Phi(p) &= RT_0 \ln\left(\frac{p_b}{p}\right) + R\Gamma (p_b-p)
\end{align*}
In the case that $p \ll p_b$, as it would if the layer under consideration is in the upper atmosphere, then the logarithmic term accounts for the majority of the variation in $\Phi$, and we can simply take
\begin{equation}
\Phi(p) = RT_0 \ln\left(\frac{p_b}{p}\right)
\end{equation}
to good approximation.
At first, we make no assumptions about the variation of flow velocity with pressure, other than separability: $\mathbf{v}(p)=\mathbf{v_0} f(p)$.  If we define $\mathbf{v}_0$ according to $\mathbf{v}(p_0)=\mathbf{v_0}$, then we must have $f(p_0)=1$.  For the time being, we place no other restrictions on $f$.
With the assumption of a suitable boundary condition, it is now possible to determine $w_p$.  As a boundary condition, we choose $w_p(0)=0$.  This gives
\begin{equation}
w_p=-\frac{F(p)}{f(p)} \nabla \cdot \mathbf{v},
\end{equation}
where
\begin{equation}
F(p) = \int_0^p f(p') \, dp'.
\end{equation}
Defining
\begin{align}
K_1 &\equiv -\frac{F(p) f'(p)}{(f(p))^2},\label{k1}\\
K_2 &\equiv \frac{F(p)}{pf(p)}\label{k2},
\end{align}
we can substitute into equations \ref{p33} and \ref{p34} to obtain a general form of the two-dimensional hydrostatically-balanced governing equations:
\begin{align}
\frac{\partial \mathbf{v}}{\partial t} =& -\mathbf{v} \cdot \nabla \mathbf{v} - K_1\mathbf{v} \nabla \cdot \mathbf{v}- R \ln \left(\frac{p_b}{p}\right) \nabla T 
%\notag \\& 
-2\Omega_{\rm{rot}} \sin \theta (\hat{n} \times \mathbf{v}) \label{p2v0}\\
\frac{\partial T}{\partial t} =& -\mathbf{v} \cdot \nabla T  - \kappa K_2 T  \nabla \cdot \mathbf{v} +f_{\rm{rad}}. \label{p2T0}
\end{align}
These equations describe atmospheric motion at a single pressure depth.  Note that isobars are in general not material surfaces; therefore, the equations do not describe the evolution of any particular set of particles in the atmosphere.  A result of this is that mass is not generally conserved, since matter is free to flow into this layer from other layers and \textit{vice versa}.  In principle, this makes the derivation of $f_{\rm{rad}}$ difficult, since the heating term is obtained by applying the first law of thermodynamics to a single parcel of gas.  However, the assumption that $\partial T/\partial p = 0$ mitigates this concern, since the material flowing into a particular zone from above or from below is expected to be at approximately the same temperature.  It is worth noting that the vertical temperature variations expected in a real atmosphere could cause as much or more error in our model due to these energy-balance concerns than due to their direct influence over the geopotential height of an isobar.  These concerns, however, are difficult to address in the context of the single-layer approach presented here.

In order to apply this model to a specific atmosphere, it is necessary to choose a specific form for $\partial \mathbf{v} / \partial p$.  If the wind shear is not large, a reasonable approximation is to take $\mathbf{v}(p)=\mathbf{v_0}(1-ap)$, so that $f(p)=1-ap$.  In this case,
\begin{align}
F(p) &= p(1-\frac{ap}{2})\\
f'(p)&=-a.
\end{align} 
Substitution into equations \ref{k1} and \ref{k2} yields
\begin{align}
K_1&=\frac{ap(1-ap/2)}{(1-ap)^2}\label{k12}\\
K_2&=\frac{1-ap/2}{1-ap}\label{k22}
\end{align}
At this point, there still seems to be little insight as to physically reasonable values for $a$.  However, examination of the pressure dependence of equation \ref{p2v0} will allow for an \textit{ad hoc} determination of the velocity shear, which should be sufficient for current purposes.  Let us assume that the scale of the velocity varies as $ln(p_b/p)$; this is tantamount to assuming that the dominant accelerating force is the temperature gradient term.  Then the velocity can be written
\begin{equation}
\mathbf{v} = \mathbf{V} \frac{\ln (p_b/p)}{\ln (p_b/p_0)}\label{Vp}.
\end{equation}
Here, $\mathbf{V}$ is the velocity at some reference pressure $p_0$, which we will momentarily identify with the $p$ that appears in equations \ref{p2v0}, \ref{k12}, and \ref{k22}.  Then
\begin{equation}
\left.\frac{\partial \mathbf{v}}{\partial p}\right|_{p_0}= -\frac{\mathbf{V}}{\ln (p_b/p_0)}\frac{1}{p_0}.
\end{equation}
A Taylor expansion of equation \ref{Vp} about $p=p_0$ therefore yields
\begin{equation}
\mathbf{v} = \mathbf{V}\left(1- \frac{p-p_0}{p_0 \ln (p_b/p_0)}\right)
\end{equation}
Rearranging,
\begin{equation}
\mathbf{v} = \mathbf{V}\left(1+\frac{1}{\ln(p_b/p_0)}\right)\left(1- \frac{p}{p_0 (1+\ln (p_b/p_0))}\right)
\end{equation}
Then we can identify 
\begin{equation}
\mathbf{v_0} = \mathbf{V}\left(1+\frac{1}{\ln(p_b/p_0)}\right)
\end{equation}
and
\begin{equation}
f(p)=1-ap=1- \frac{p}{p_0 (1+\ln (p_b/p_0))}\label{feq}
\end{equation}
We are interested in the value of $ap$ at the layer under consideration -- that is, at $p \rightarrow p_0$ in the Taylor expansion.  Making this substitution in equation \ref{feq} yields the desired value of $ap$:
\begin{equation}
ap=\frac{1}{1+\ln (p_b/p)},
\end{equation}
where we have dropped the subscript from $p_0$.
Defining $\alpha_1\equiv \ln (p_b/p)$ and $\alpha_2 \equiv 1/(1+\alpha_1)$, we can write the final form of the two-dimensional hydrostatically-balanced governing equations for an atmospheric layer at constant pressure:
\begin{align}
\frac{\partial \mathbf{v}}{\partial t} &= -\mathbf{v} \cdot \nabla \mathbf{v} - \left(\frac{\alpha_2(1-\alpha_2/2)}{(1-\alpha_2)^2}\right)\mathbf{v} \nabla \cdot \mathbf{v}- R \alpha_1 \nabla T -2\Omega_{\rm{rot}} \sin \theta (\hat{n} \times \mathbf{v}) \label{p2v}\\
\frac{\partial T}{\partial t} &= -\mathbf{v} \cdot \nabla T  - \kappa \left(\frac{1-\alpha_2/2}{1-\alpha_2}\right)T  \nabla \cdot \mathbf{v} +f_{\rm{rad}}. \label{p2T}
\end{align}
These equations govern the time-evolution of $T$ and $\mathbf{v}$, and, when augmented by an appropriate treatment of the radiative forcing term $f_{\rm{rad}}$, are suitable for atmospheric simulation.  

To derive the radiative forcing term, we assume that the layer under consideration extends from some pressure $p,$ which corresponds to the infrared photosphere, to arbitrarily low pressures.  We assume that this layer absorbs some fraction $X$ of incident stellar flux.  While $X$ is effectively a free parameter in the model, in principle it can be determined based on atmospheric chemistry and appropriate $p$-$T$ profiles.  The remaining portion of incident radiation is absorbed deeper in the atmosphere, where it is reradiated through thermal emission.  This is modeled by an isothermal layer which always remains at some constant $T_{n}$, where $T_{n}$ is a function of both the amount of penetrating solar radiation and internal heating due to tidal dissipation, gravitational contraction, etc.   The assumption here is that below the infrared photosphere, advective processes quickly redistribute heat and destroy any temperature variation.  This is, of course, a considerable oversimplification of the real atmosphere, but it will have to suffice for our purposes.  It is also assumed that all of the thermal emission from this interior layer is absorbed by the upper layer being simulated.  Should it prove necessary to relax this second assumption -- for example, to model extremely low pressure depths -- the adjustments to accomplish this are straightforward.

In this model, the energy input from the absorption of incident sunlight is simply $XF_*(0)$.  The energy input due to absorption of thermal emission from deeper in the atmosphere is simply $\sigma T_{n}^4$, since the outgoing long-wave radiation from the internal layer is completely absorbed.  $T_{n}$ is calculated by time-averaged energy balance, wherein the stellar flux penetrating to the deeper layer is averaged over the the course of an orbit.  The time-averaged penetrating flux is
\begin{equation}
F_{\rm{pen}} = (1-A)(1-X)\left(\frac{L_*}{16 \pi a^2\sqrt{1-e^2}}\right),
\end{equation}
where the additional factor of $\sqrt{1-e^2}$ in the denominator arises from the time-averaging, and represents the disproportionate contribution to the heating during the periastron passage.

In addition to the heating of the internal layer produced by stellar irradiation penetrating the upper layer, the planet itself can provide a considerable contribution to the energy budget, due to tidal heating, gravitational contraction, and possibly other effects.  These are encapsulated in the $T_{\rm{int}}$ parameter, which is related to the planet's intrinsic luminosity by $L_{\rm{int}}=\sigma T_{\rm{int}}^4$.
With these two sources of heating, energy balance requires that the inner layer maintain a steady-state temperature
\begin{equation}
T_{n}=\left(\frac{F_{\rm{pen}}}{\sigma}+T_{\rm{int}}^4\right)^{1/4}. \label{Tneq}
\end{equation}

Energy balance for the upper layer therefore requires that
\begin{equation}
\sigma_c c_p f_{\rm{rad}}  = XF_*(0) + \sigma T_{n}^4 - \sigma T^4,
\end{equation}
where $\sigma_c=p/g$ is the column density of the upper layer and $p$ is the pressure at the $bottom$ of the upper layer.

Rearranging and explicitly writing out the contribution from $F_0$, we have
\begin{equation}
f_{\rm{rad}}= \left( \frac{\sigma g}{p c_p} \right)\left(X(1-A)\left(\frac{L_*}{4 \pi \sigma a^2}\right) \cos \alpha+T_{n}^4-T^4 \right). \label{rad2}
\end{equation}

This model cannot, of course, provide either the realism or the physical insight that is achievable within the framework of genuine radiative transfer, which explicitly accounts for both the vertical structure and the chemistry of the atmosphere.  However, it offers significantly greater accuracy than Newtonian heating \citep{cho03, coo05, cho08, sho08}, particularly in the case that the temperature strays far from its equilibrium value.  Unlike a full treatment of radiative transfer, it imposes negligible additional computational cost.

The hydrodynamic equations \ref{p2v} and \ref{p2T}, combined with the radiative forcing described by equations \ref{Tneq} and \ref{rad2} provide a set of equations which are suitable for numerical integration.  As in our previous simulations \citep{lan07, lan08, lan08b}, the numerical integration is accomplished using the vector spherical harmonic transform procedure described in \citet{ada99}.  To obtain a numerical solution, specific values for three parameters must be chosen: the base pressure $p_b$, the pressure at the simulated layer $p$, and the fraction $X$ of absorbed incident light.

The dynamics are not particularly sensitive to the choice of $p_b$; we take $p_b=4$ bar, which is a reasonable value.  Larger values tend to increase wind speed, but due to the fact that the winds are coupled only logarithmically to $p_b$, the effect is fairly small.
The choice of $p$, the pressure at the simulated layer, determines both the wind speed and the radiative time-scale.  The pressure depth of the 8-$\mu$m photosphere cannot be determined precisely without a fairly detailed understanding of the atmospheric chemistry and structure; we therefore assume $p$ to be a free parameter, albeit one which is relatively constrained by physical considerations.  The results presented here are based upon simulations with 100 mbar $\le p \le$ 400 mbar.

Likewise, the fraction $X$ of incident sunlight absorbed in the simulated layer must also be treated as a free parameter.  However, $X$ and $T_{n}$ are fairly well-constrained in the case of HD 189733 b by the observed flux variation.  While a full parameter study might yield values here which produce more precise fits, we find that taking $X=0.5$, leading to $T_{n} = 975 K$, gives sufficient accuracy for our purposes in this preliminary investigation.  We note here that $T_{n}$ depends upon both $X$ and the intrinsic temperature $T_{\rm{int}}$, which is produced by internal heating.  For this planet, we take a conservative value of $T_{\rm{int}}=100$ K.  For such low values of the internal heating, $T_{n}$ is determined almost completely by the reabsorption of thermal emission from deeper atmospheric layers.

%\clearpage
\section{Results}

To lowest order, our simulated flows agree with those obtained by \citet{sho08}.  In the uppermost portions of the atmosphere ($p\lesssim100-150$ mbar), air flows from the substellar point towards the antistellar point, so that the flow is westward on the evening terminator and eastward on the morning terminator.  At deeper pressures, however, a persistent eastward equatorial jet develops, with a characteristic wind speed of $\mathbf{v} \approx 3.5$ km/s.  (At the pressure depths considered in these simulations, the maximum wind speed typically falls in the range 3.5~km/s~$\le \mathbf{v} \le$~5~km/s.)  This jet is supersonic: the speed of sound on HD 189733 b varies between 2.2 km/s in the coldest regions and 3 km/s in the hottest. 

\begin{figure}
\begin{center}
\includegraphics{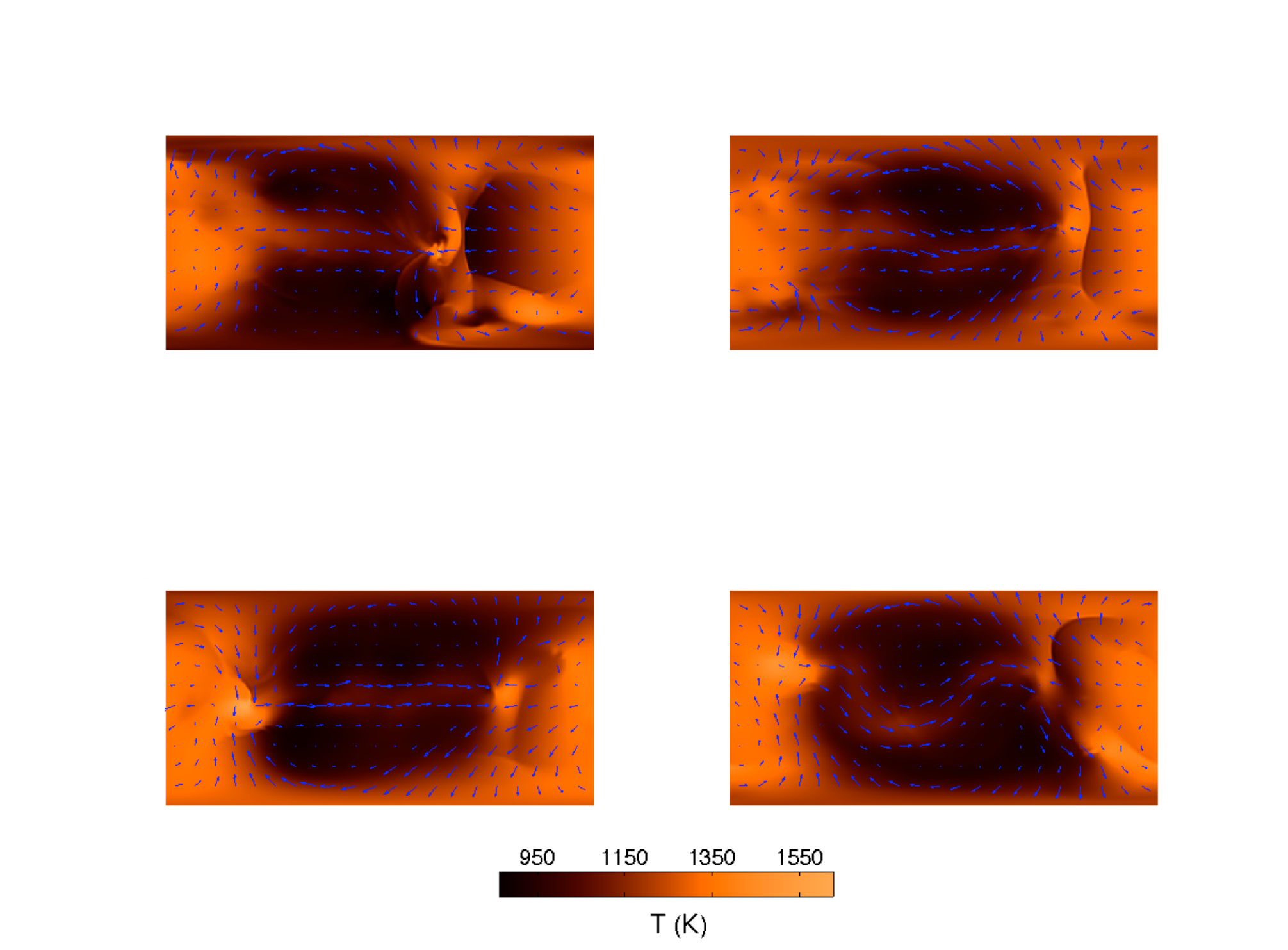}
\end{center}
\caption{Evolution of the flow pattern and temperature distribution in the atmosphere of HD 189733 b over two orbits.  The top left panel is at $t=0$, the top right at $t=2P/3$, the bottom left at $t=4P/3$, and the bottom right at $t=2P$.  The substellar point on these equirectangular plots is at the center of the leftmost (or rightmost!) edge, while the antistellar point is directly in the center of each plot.  These flows were generated assuming $X=0.5$ and $p=150$ mbar; under these assumptions, the equatorial jet reaches speeds in excess of 4 km/s.}
\label{189733flow}
\end{figure}

A key difference between the atmospheric flow produced by our model and that found by \citet{sho08} is the development of large-scale turbulence in our simulations; the flow in the \citeauthor{sho08} model is laminar.  This is a robust characteristic of our simulations: some degree of turbulence arises at all pressure depths considered.  As might be expected, it is most extreme at lower pressures, where the wind speed is greater.  In general, one would expect to find turbulent flow in the upper atmosphere, due to the tremendous wind shears generated by the head-on collision of eastward and westward flows that characterize the lowest pressures.  The absence of turbulent conditions in the \citeauthor{sho08} results is therefore perhaps surprising.  The evolution of the flow can be seen in figure \ref{189733flow}, which shows four equirectangular plots showing the wind velocity superimposed on the temperature distribtion.

\begin{figure}
\begin{center}
\includegraphics{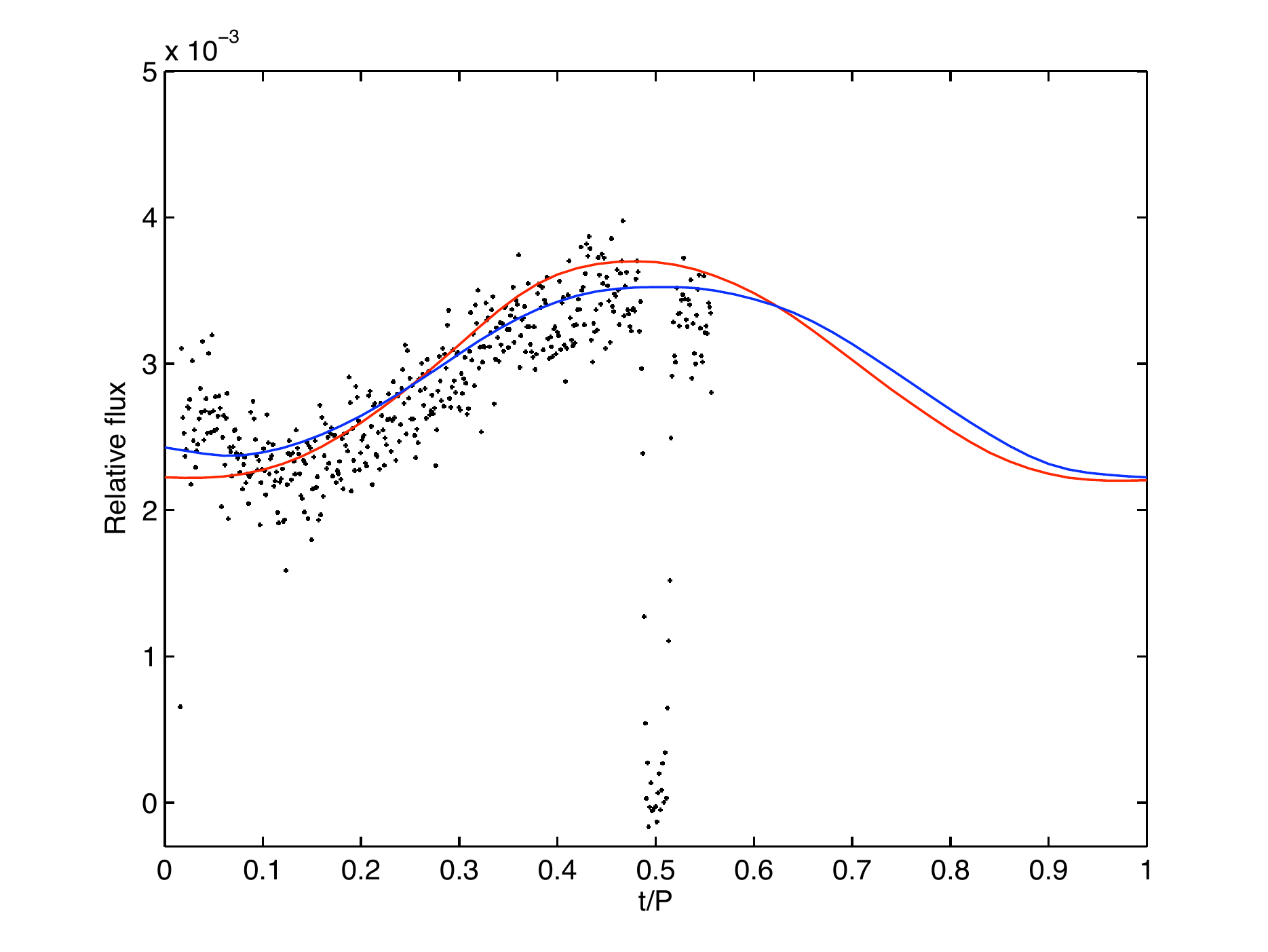}
\end{center}
\caption{Predicted infrared flux at 8 $\mu$m from HD 189733 b on two successive orbits.  The variation between orbits is likely to be significant if the model used here is correct.  Transit corresponds to $t/P=0$ and $t/P=1$, while the secondary eclipse occurs at $t/P=0.5$.During the first orbit -- the blue curve -- the flux minimum occurs 3.6 hours after (the center of) transit, while the flux maximum occurs 0.36 hours after secondary eclipse.  During the second orbit, shown in red, the flux minimum occurs 0.81 hours after transit, with the maximum coming 1.1 hours before the secondary eclipse.  The black points show the data obtained after transit by \citet{knu07}.}
\label{189733lc}
\end{figure}

\begin{figure}
\begin{center}
\includegraphics{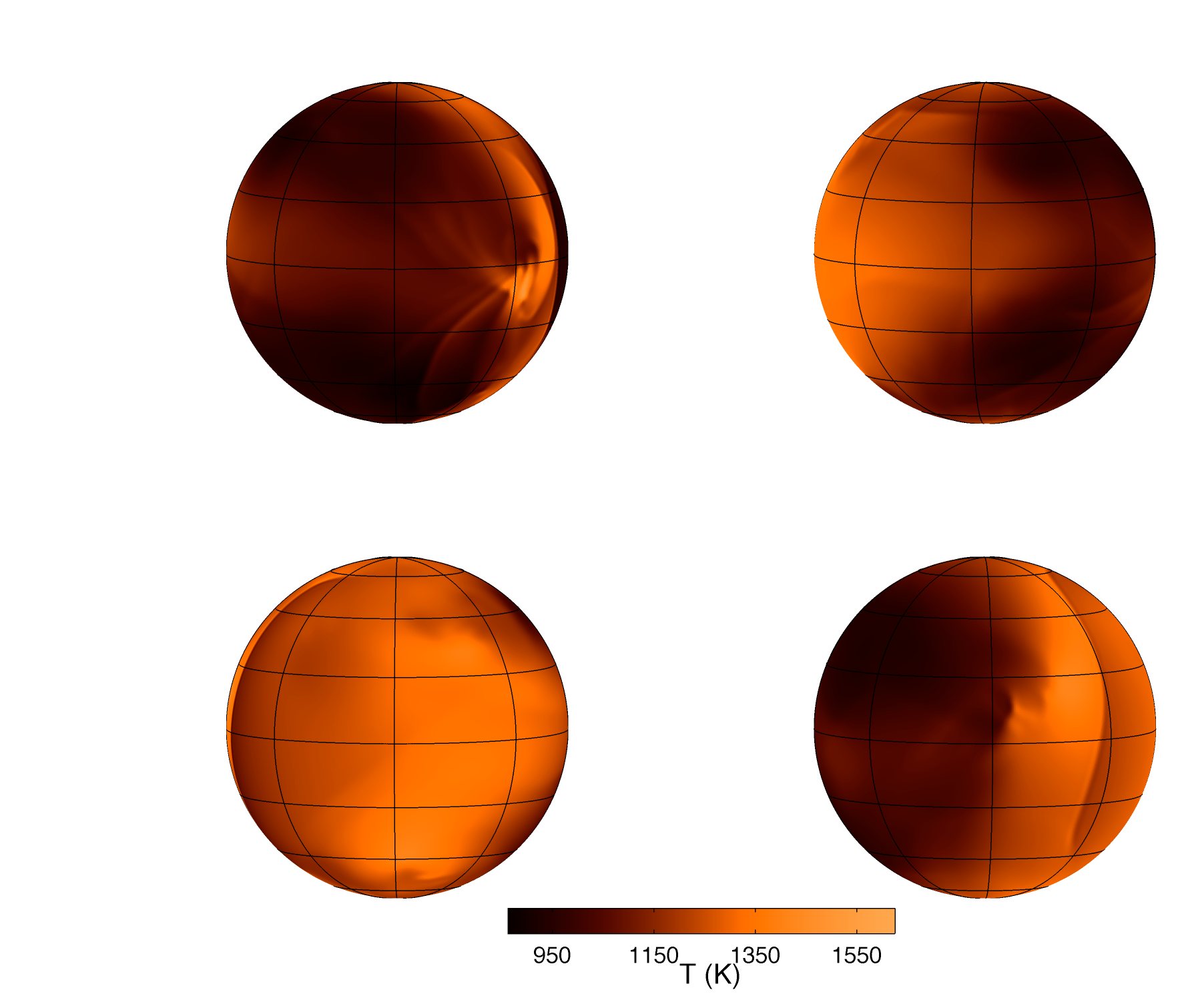}
\end{center}
\caption{Evolution of the temperature distribution at the $p=150$ mbar layer of HD 189733~b over one orbital period, as seen from Earth.  The four globes correspond to the following times, measured from transit: top left, $t=0$; top right, $t=P/4$; bottom left, $t=P/2$; bottom right, $t=3P/4$.  Hemispherical integration of these temperature distributions leads to the light curve shown blue in figure \ref{189733lc}.}
\label{189733earth1}
\end{figure}

\begin{figure}
\begin{center}
\includegraphics{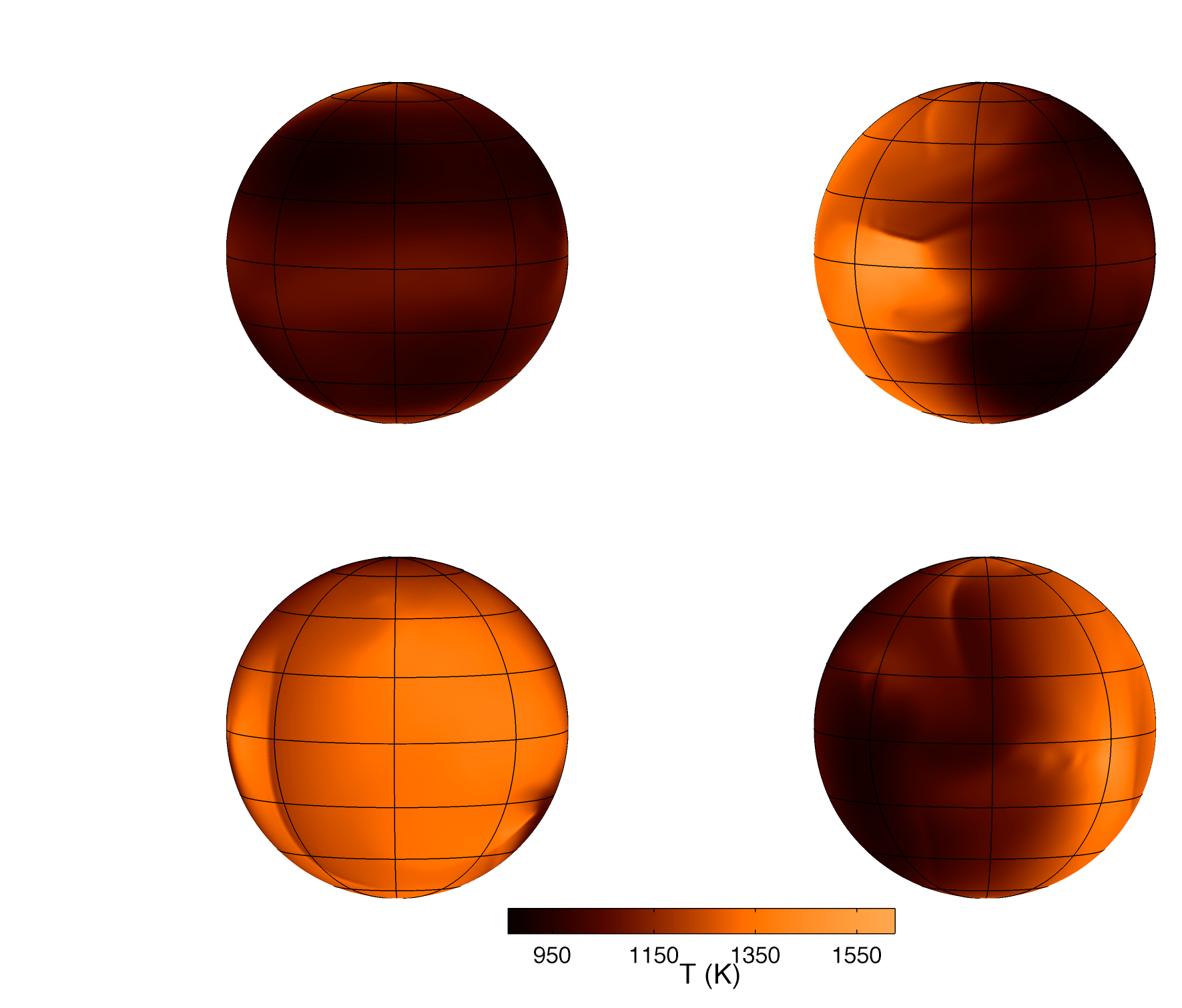}
\end{center}
\caption{Evolution of the temperature distribution at the $p=150$ mbar layer of HD 189733~b over a second orbital period, as seen from Earth.  The four globes correspond to the following times, measured from transit: top left, $t=0$; top right, $t=P/4$; bottom left, $t=P/2$; bottom right, $t=3P/4$.  Hemispherical integration of these temperature distributions leads to the red curve shown in figure \ref{189733lc}.}
\label{189733earth2}
\end{figure}

The flux variation depends primarily on the choice of $X$, only weakly varying with $p$.  For $X=0.5,$ we find a maximum relative flux $F_{\rm{max}}/F_* \approx 3.6 \times 10^{-3}$, with a minimum relative flux of $F_{\rm{min}}/F_* \approx 2.2 \times 10^{-3}$, with variations between orbits on the order of $2\times10^{-4}$.  This is in good agreement with \citet{knu07}, who find a total flux variation $F_{\rm{max}}/F_*-F_{\rm{min}}/F_* = 1.2\times10^{-3} \pm 0.2\times10^{-3}$, with $F_{\rm{min}}/F_{\rm{max}} = 64 \% \pm 7 \%$: the results of our model yield $F_{\rm{max}}/F_*-F_{\rm{min}}/F_* = 1.4\times10^{-3}$, with $F_{\rm{min}}/F_{\rm{max}} = 61\%$.

Our model shows significant variation in the shape of the light-curve from one orbit to the next, as can be seen in figure \ref{189733lc}.  While the overall flux variation changes by only a small amount, the phase offset of the flux extrema from the time of central transit is measurably different from one orbit to the next.  In the first orbit shown in figure \ref{189733lc}, the flux minimum occurs 3.6 hours after (the center of) transit, while the flux maximum occurs 0.36 hours after secondary eclipse.  During the second orbit, shown in red, the flux minimum occurs 0.81 hours after transit, with the maximum coming 1.1 hours before the secondary eclipse.  The evolving temperature distribution that gives rise to these light-curves is shown in figures \ref{189733earth1} (first orbit; blue curve in figure \ref{189733lc}) and \ref{189733earth2} (second orbit; red curve).

While it is currently impractical rigorously to determine the positions of the flux minima and maxima over many orbits, it is possible to estimate the range over which they can be expected to vary.  We take a ``snapshot'' of the temperature distribution twice per orbital period over thirty orbits following the establishment of a dynamical equilibrium.  We then determine the light-curve that would result if the temperature distribution were to remain constant over the course of an entire period.  This provides a zeroth-order approximation to the variation in the shapes of light-curves that might be expected if one were to observe many orbits.  From the sixty light-curves that are obtained through this method, the mean position of the flux maximum was 0.89 hours before the secondary eclipse, with results ranging from 2.6 hours prior to the eclipse to 1.1 hours after the eclipse.  The mean position of the flux minimum was 0.69 hours after transit, with results ranging from 4.2 hours before transit to 4.7 hours after transit.  In future work, a more rigorous determination of these variations will be necessary.  It is also important to note that the times for flux maximum and minimum seem to depend rather strongly on the pressure depth of the simulation, with the largest variations occurring in the highly turbulent region between 100 mbar and 200 mbar.

These results are qualitatively consistent with the shape of the light-curve determined by \citet{knu07}.  In the \citeauthor{knu07} data, the flux minimum is found to occur $6.7 \pm 0.4$ hours after the transit, with the flux maximum preceding the secondary eclipse by $2.3 \pm 0.8$ hours.

It is worthwhile here to note that turbulent shallow-water simulations performed by \citet{rau07} also yield potentially variable light curves.  The dynamics in the \citeauthor{rau07} model are completely different from those posted here: wind speeds are limited to $\lesssim 800$ m/s, and the flux variation is produced by the motion of cold spots produced by persistent circumpolar vortices, rather than the more chaotic collision of supersonic jets seen in our model.  Nevertheless, it is instructive that those models in which the flow is turbulent produce variable light curves, while the light curves resulting from the laminar flows seen in \citet{coo05} and \citet{sho08} appear to be essentially constant from one orbit to the next.

\begin{figure}
\begin{center}
\includegraphics{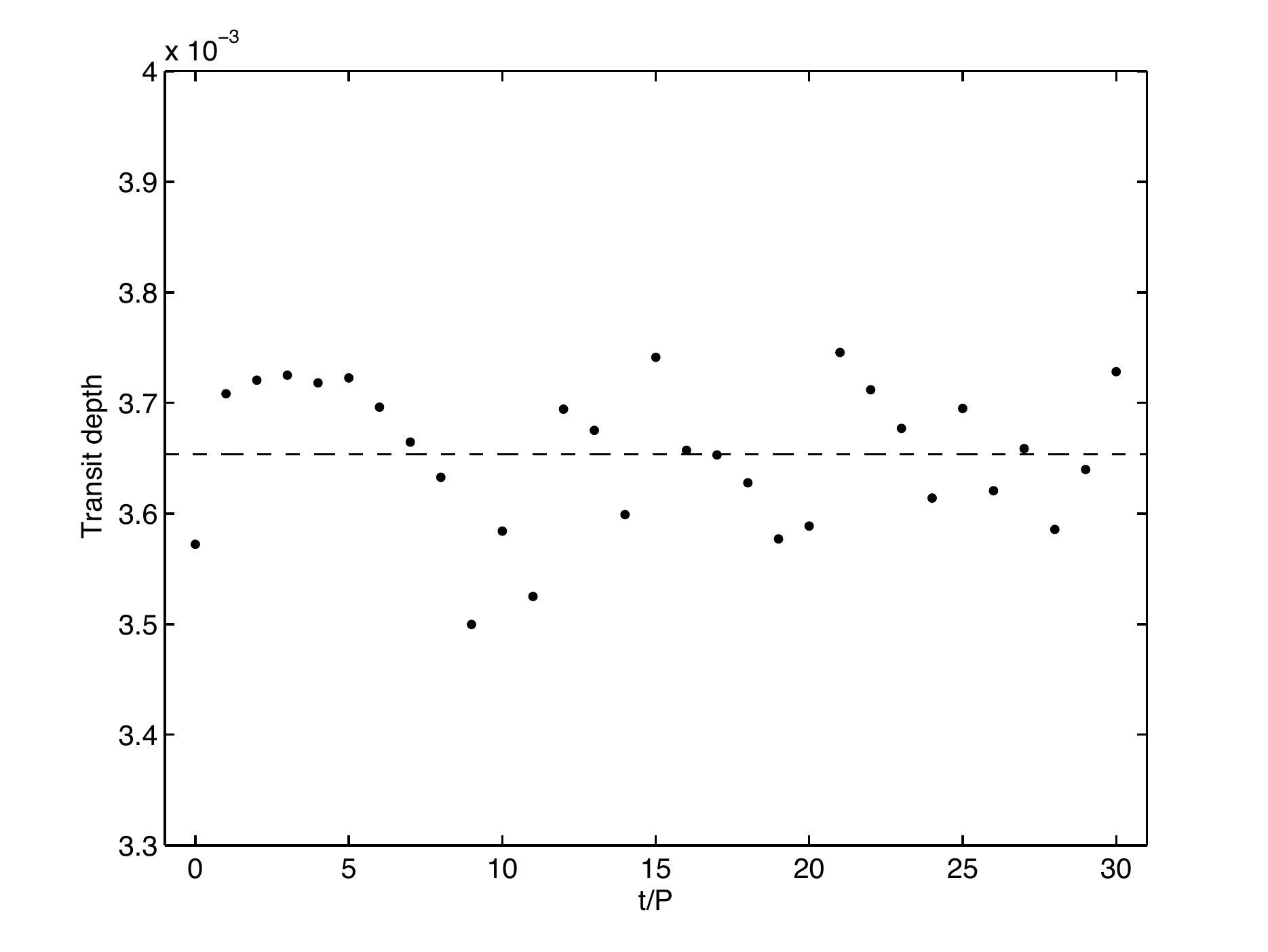}
\end{center}
\caption{Model prediction for variability in the depth of secondary transit at 8 $\mu$m.  The points show the transit depth over thirty successive orbits, while the dashed line shows the average transit depth of $0.365\%$.}
\label{trns}
\end{figure}

Recent observations of the the depth of secondary transit over several orbits by \citet{ago08} place further constraints on the allowable inter-orbit variations.  \citeauthor{ago08} find less than about $10\%$ deviation from their mean observed transit depth of $0.347\%$.  In figure \ref{trns}, we show the secondary transit depth over 30 successive orbits, as predicted by our model.  We obtain a slightly larger mean transit depth of $0.365\%$, with a maximum variation of about $5\%$.  Thus, the model appears to be fully consistent with the \citeauthor{ago08} observations.

\section{Conclusions}
While this model has not yet provided an optimized fit to the light-curve of HD 189733 b obtained by \citet{knu07}, it does provide a reasonable explanation for a dynamical configuration which could give rise to the observed light curve.  In general, other models have been able to reproduce the observed phase offset of the flux maximum, corresponding to an eastward shift of the hottest temperatures from the substellar point \citep{coo05, sho08}.  It has been more difficult to account for the apparent westward shift of the coldest temperatures from the anti-stellar point.  However, the model proposed in this paper offers a physically reasonable mechanism by which the observed light curve may arise: at low pressures $p<200$ mbar, the flow tends to run from the substellar point to the antistellar point at supersonic speeds.  This results in a collision between eastward winds and westward winds, which produces chaotic, turbulent flow on a large scale.  As a result, there is no steady-state temperature distribution; the enormous winds cause unpredictable shifts in the temperature distribution which, in turn, alter the position of the flux minima and maxima from orbit to orbit.  The phase offset of the flux minimum observed by \citet{knu07} is not inconsistent with these results.

Furthermore, this hypothesis is readily testable: turbulence on the scale predicted by our model is expected to induce shifts of several hours or more in the timing of the flux minima and flux maxima; the light-curve obtained by \citet{knu07} indicates that it is possible to measure the timing of these flux extrema with an error of $\lesssim \pm 1$ hour.  Therefore, if subsequent observations of HD 189733 b showed significant variations in the infrared light curve from one orbit to the next, the presence of significant large-scale turbulence would be strongly supported.  Conversely, the absence of such variations would imply that turbulence on a scale necessary to explain the \citeauthor{knu07} results is suppressed by other factors for which this simple two-dimensional model is unable to account.  Some support for the model presented here may be found from the recent 24-$\mu$m phase curve obtained by \citet{knu08}.  These observations were taken over the same portion of the orbit as the earlier 8-$\mu$m observations \citep{knu07}.  In both cases, the flux maximum precedes secondary eclipse.  In contrast to the 8-$mu$m data, however, the 24-$\mu$m curve increases monotonically throughout the observing window -- there is no flux minimum following the transit.  It is unclear whether this is due to a qualitatively different flow at the 24-$\mu$m photosphere, or due to the type of turbulent variation suggested by the model presented in this paper.  It would be useful to obtain more observations in both the 8- and 24-$\mu$m bands so that a comparison can be made between flows on multiple orbits, but at the same atmospheric depth.   In any case, it is quite clear that extensive further observations are necessary to obtain a detailed characterization of the atmospheric behavior of HD 189733 b.

The model presented here represents a significant forward step from previous two-dimensional models.  Although the number of free parameters in our treatment of the radiative forcing mitigates our success in fitting the \citet{knu07} light curve, the quality of the fit is sufficient to conclude that the dynamics produced by our model offer a reasonable explanation for the observations, while a poor fit would imply that the simulated flows are excluded by the data.  We can therefore say with some confidence that our model is not excluded by the observations currently available.  It is also the first model -- in either two or three dimensions -- to provide an explanation for the unexpected phase offset of the flux minimum in the HD 189733 b 8-$\mu$m curve.

Despite these encouraging results, much room for improvement exists.  A more sophisticated treatment of the radiative forcing is necessary.  However, true radiative transfer is difficult to approximate in a two-dimensional model; among other issues, the depth to which incident radiation penetrates depends rather strongly on the wavelength.  It is therefore likely that improvements in this area will require a shift to a fully three-dimensional code.  Furthermore, a three-dimensional treatment is necessary to ensure that departures from stratified flow do not significantly affect the results.  However, the ubiquity of turbulent flow obtained in two-dimensional simulations suggests that small-scale motion is non-negligible, so that the low horizontal resolutions seen current three-dimensional models are less than ideal.  An immediate goal is therefore the production of a high-resolution three-dimensional model, including a realistic multi-wavelength treatment of radiative transfer.

With the recent decision to fund a non-cryogenic \textit{Spitzer} mission motivated at least in part by the promise of useful observations of exoplanets, the efforts of modelers to assist in the identification of interesting targets will be critical.  Whatever uncertainties still plague these efforts, it is certain that the influx of \textit{Spitzer} data over the next few years should provide rigorous tests on both existing models and on those yet to be developed.

\acknowledgments

We are grateful to Jonathan Fortney for useful suggestions.  The code employed in this paper makes use of the SPHEREPACK 3.0 routines written by Drs. Adams and Swarztrauber and provided by NCAR.  This research has been supported
by the NSF through CAREER Grant AST-0449986, and by the NASA Planetary Geology and Geophysics Program through Grant NNG04GK19G.

\end{document}